\documentclass{article}
\usepackage{graphicx} % Required for inserting images
\usepackage[a4paper,left=2cm,right=2cm,top=2cm,bottom=2cm]{geometry}
\usepackage{tikz}
\usepackage{authblk}
\usetikzlibrary{shapes,arrows}
\usetikzlibrary{positioning,chains}
\usepackage{url}
\usepackage{booktabs}
\usepackage{multirow}
\usepackage{lscape}
\usepackage{longtable}
\usepackage{tabularray}
\usepackage{booktabs,caption}
\usepackage[flushleft]{threeparttable}
\usepackage{lipsum}
\usepackage{threeparttable}
\usepackage{booktabs}
\usepackage{dcolumn}
\usepackage{pdflscape}
\usepackage{longtable}
\usepackage{threeparttablex}
\usepackage{setspace}
\usepackage{lineno}
\usepackage{lscape}
\usepackage{fancyhdr}
\title{USE OF SURROGATE ENDPOINTS IN HEALTH TECHNOLOGY ASSESSMENT: A REVIEW OF SELECTED NICE TECHNOLOGY APPRAISALS IN ONCOLOGY}

\author[1,*]{Lorna Wheaton, MSc}
\author[1]{Sylwia Bujkiewicz, PhD}

\date{}

\affil[1]{Biostatistics Research Group, Department of Population Health Sciences, University of Leicester, University Road, Leicester, LE1 7RH, UK}
\affil[*]{Corresponding author: Lorna Wheaton, lskw3@leicester.ac.uk}

\newpage

\begin{document}

\maketitle

\newpage

\section*{Abstract}
\noindent
\textbf{Objectives:} Surrogate endpoints, used to substitute for and predict final clinical outcomes, are increasingly being used to support submissions to health technology assessment agencies. The increase in use of surrogate endpoints has been accompanied by literature describing frameworks and statistical methods to ensure their robust validation. The aim of this review was to assess how surrogate endpoints have recently been used in oncology technology appraisals by the National Institute for Health and Care Excellence (NICE) in England and Wales.

\noindent
\textbf{Methods:} This paper identifies technology appraisals in oncology published by NICE between February 2022 and May 2023. Data are extracted on the use and validation of surrogate endpoints including purpose, evidence base and methods used. 

\noindent
\textbf{Results:} Of the 47 technology appraisals in oncology available for review, 18 (38 percent) utilised surrogate endpoints, with 37 separate surrogate endpoints being discussed. However, the evidence supporting the validity of the surrogate relationship varied significantly across putative surrogate relationships with 11 providing RCT evidence, 7 providing evidence from observational studies, 12 based on clinical opinion and 7 providing no evidence for the use of surrogate endpoints. 

\noindent
\textbf{Conclusions:} This review supports the assertion that surrogate endpoints are frequently used in oncology technology appraisals in England and Wales and despite increasing availability of statistical methods and guidance on appropriate validation of surrogate endpoints, this review highlights that use and validation of surrogate endpoints can vary between technology appraisals which can lead to uncertainty in decision-making. 

\noindent
\textbf{Keywords:} surrogate endpoint; health technology assessment; decision making

\section{Introduction}

Surrogate endpoints have become an integral part of the development of novel health technologies and in particular in oncology. Before a new cancer therapy can reach patients, it must first be approved by regulatory agencies such as the European Medicines Agency (EMA) in the European Union or the Food and Drug Administration (FDA) in the United States. Traditionally, regulatory approvals were based on data from randomised controlled trials (RCTs) demonstrating the treatment effect of the new health technology on overall survival (OS), which was considered the most objective and clinically meaningful outcome of most importance to patients as well as policy makers. However, increasingly regulatory agencies have been carrying out accelerated approvals of health technologies based on treatment effects observed on surrogate endpoints. Surrogate endpoints are intermediate outcomes which can substitute for and predict final clinical outcomes of interest such as mortality~\cite{ciani2016use, ciani2017time, ciani2022development, demets2020can, gyawali2019assessment, robb2016biomarkers}. The use of surrogate endpoints as primary outcomes in clinical trials can result in smaller, cheaper and shorter studies compared to trials which are designed to evaluate treatment effectiveness on the final clinical outcome such as OS. 

Although the use of surrogate endpoints accelerates the evaluation of therapies in clinical trials and regulatory approvals, in many countries patient access to the new treatment also depends on subsequent scrutiny by health technology assessment (HTA) agencies such as the National Institute for Health and Care Excellence (NICE) who provide reimbursement recommendations for England and Wales. Whilst regulatory decisions are based on the safety and efficacy of the product, HTA agencies are primarily concerned with the long-term clinical and cost-effectiveness of the new health technology in order to ensure that scarce healthcare resources are used efficiently. Therefore, the increasing reliance on surrogate endpoints for regulatory approvals creates problems for HTA agencies as long-term estimates of clinical and cost-effectiveness are unavailable at the time of HTA submissions that closely follow regulatory approvals. As such, HTA agencies generally express a strong preference for final clinical outcomes over surrogate endpoints and where these are not available require that surrogate endpoints are appropriately validated. 

Over recent years there has been a wealth of literature published regarding the prolific use of surrogate endpoints in HTA and in particular in oncology indications~\cite{kemp2017surrogate, dawoud2021raising, ciani2021validity}. There has also been a significant amount of criticism regarding such use of surrogate endpoints and calling for more transparent frameworks and harmonisation of guidelines across HTA agencies~\cite{ciani2021validity, dawoud2021raising, grigore2020surrogate}. 
A plethora of statistical methods have been developed to ensure robust evaluation of surrogate endpoints~\cite{ensor2016statistical} including NICE Decision Support Unit's technical support document 20 (TSD20) which describes methods for multivariate meta-analysis to evaluate surrogate endpoints~\cite{bujkiewicz2019nice}. NICE have also updated their methodology guide to ensure improved evaluation of surrogate endpoints in submissions to their technology appraisals (TAs). 
Whilst available meta-analytic methods  serve well the purpose of surrogate endpoint validation and predicting the relative treatment effect on the final clinical outcome (which can be used in decision modelling frameworks), they do not capture all aspects of the role a surrogate endpoint can play in the decision modelling. 
A methodological review, carried out by the NICE Decision Support Unit in preparation for the update of NICE methodology guide, recommended a review of past technology appraisals to fully understand how surrogate endpoints have been used in decision making by NICE to inform further methodological considerations and identify most optimal modeling approaches utilising surrogate endpoints~\cite{welton2020chte2020}.
With the context of such recent developments in the field of surrogate endpoints in the HTA setting, the aim of this project was to review and assess how surrogate endpoints have recently been used in NICE oncology technology appraisals. Specifically we are interested in how often surrogate endpoints have been used, the justification for their use, the level of evidence provided to support the use of the surrogate endpoint and in what way they informed health economic modelling.

\section{Methods}

Documents relating to NICE TAs are published on the NICE website. We retrieved documents from all NICE TAs for cancer drugs evaluated between 1st February 2022 and 3rd May 2023 from the NICE Guidelines website. The date of 1st February 2022 was chosen to coincide with the NICE methods guide update which included updated guidelines on how surrogate endpoints should be used in NICE technology appraisals~\cite{NICEupdates}. Terminated, withdrawn or replaced appraisals were removed as documentation was unavailable. A quality appraisal of the extracted technology appraisals was not conducted as every technology appraisal conducted by NICE is subject to review by an evidence review group (ERG) who assess the clinical and cost effectiveness analyses submitted by the company. An overview of the workflow can be seen in Figure \ref{fig:flowchart}. 

In the first stage, we adapted computer script from Polak et al~\cite{polak2022real} to automatically list and download all documentation (e.g. manufacturer submissions, evidence review group (ERG) reports and final appraisal determinations) available through the NICE website. For all TAs with documents available, the following general information was extracted by LW from either the company submission, ERG report or final appraisal decision documents:

\begin{itemize}
    \item Appraisal ID
    \item Appraisal date
    \item Treatment
    \item Disease area
    \item Primary endpoint(s) of pivotal trial
    \item Secondary endpoint(s) of pivotal trial
    \item Recommendation
\end{itemize}

In the second stage, these documents were screened by LW to assess whether they contained either of the terms ``surrogate" or ``surrogacy". If either of these terms were present, the context of the term was reviewed to make an initial assessment of whether surrogate endpoints were utilised in either the clinical or cost-effectiveness analysis.

Following this initial assessment, in the third and final stage, the clinical and cost-effectiveness sections of the committee papers were reviewed to further assess whether a surrogate endpoint was utilised. For studies which included discussion of surrogate endpoints, it was important to understand not only which surrogate endpoints were used, but also if the surrogate endpoint was validated. Validation of a surrogate endpoint requires evidence from the literature that achieving a treatment effect on the surrogate endpoint will reliably predict a clinically meaningful treatment effect on the final clinical outcome~\cite{fleming2012biomarkers}. Ciani et al~\cite{ciani2017time} suggest a three stage approach for validating surrogate endpoints for use in health care decision making. The first stage is to establish the level of evidence available for the surrogate. The authors suggest that there are three levels of evidence where the lowest level is biological plausibility of the surrogate relationship, the second level is an association between the surrogate endpoint and final outcome and the highest level is where the treatment effect on the surrogate endpoint is associated with the treatment effect on the final outcome. Once the level of evidence is established, the second step is to assess the strength of association between the surrogate endpoint and final outcome and the final step is to quantify the relationship between the treatment effects on the surrogate endpoint and the final outcome. Based on this framework, the following information was extracted from either the company submission or ERG report: 

\begin{itemize}
    \item what was the surrogate endpoint
    \item what was the final clinical outcome 
    \item what was the level of evidence for surrogate relationship 
    \item whether association between surrogate endpoint and final outcome was investigated and method used or measure of such association
    \item whether association between treatment effects on the surrogate endpoint and on the final outcome was investigated and method or measure of such association used
    \item whether predictions of the treatment effect on the final outcome from the effect on the surrogate endpoint were considered, including methods used
\end{itemize}

We also assessed whether and how information on the surrogate endpoint was utilised in the health economic evaluation.

\section{Results}

A total of 65 technology appraisals (TAs) in oncology were published on the NICE website between 1st February 2022 and 3rd May 2023. Of these 65 TAs, 16 appraisals did not provide a submission and 2 appraisals did not provide committee papers for review. The remaining 47 appraisals had 663 documents which were downloaded and screened. 

94 percent of appraisals (44/47), included at least one mention of the terms ``surrogate" or ``surrogacy". However, initial assessment of the context of the terms found that for 36 percent of appraisals (16/44), there was no meaningful discussion of surrogate endpoints. For example, in 15 TAs, the word ``surrogate" was included as part of a question in the clinical expert statement or professional organisation submission asking ``if surrogate outcome measures were used, do they adequately predict long-term clinical outcomes". While the majority of these TAs did not provide a response to this question, several included surrogacy terms by making statements such as ``no surrogate outcome measures were used" (TA819)~\cite{TA819}. 

The exclusion of three TAs for failing to contain any terms relating to surrogacy and a further 16 TAs for failing to contain any meaningful discussion of surrogate endpoints resulted in 28 TAs for a more in depth review. Upon reviewing the clinical and cost-effectiveness sections of the committee papers for these TAs, a further 10 appraisals were excluded for not including surrogate endpoints in the technology assessment. This left 18 appraisals which discussed use of surrogate endpoints and were included for full data extraction.

\begin{center}
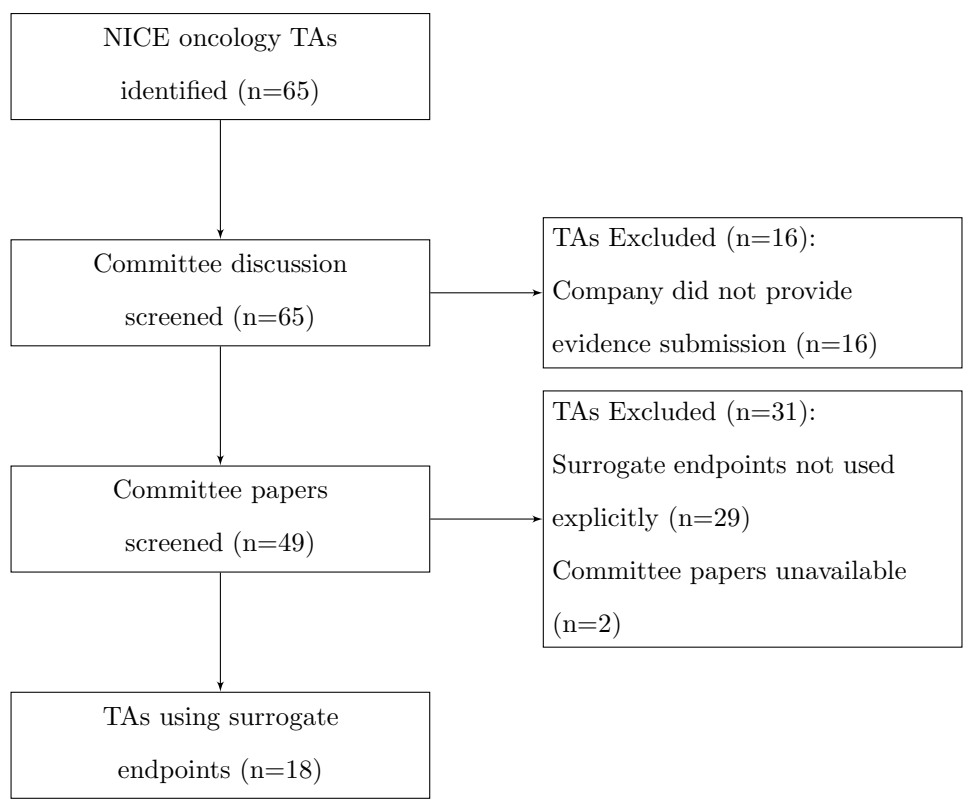

% Define block styles
\tikzstyle{block1} = [rectangle, draw, text width=15em, text centered, minimum height=4em]
\tikzstyle{block2} = [rectangle, draw, text width=15em, text ragged, minimum height=4em]
\tikzstyle{line} = [draw, -latex']
    
\begin{tikzpicture}[node distance = 2cm, auto]
    % Place nodes 
    \node[block1] (records) {NICE oncology TAs identified (n=65)};
    \node[block1, below of=records, node distance=3cm] (titles) {Committee discussion \\ screened (n=65)};
    \node[block2, right of=titles, node distance=7cm] (excluded1) {TAs Excluded (n=16):\\ Company did not provide \\evidence submission (n=16)};
    \node[block1, below of=titles, node distance=3cm] (fulltext) {Committee papers screened (n=49)}; 
    \node[block2, right of=fulltext, node distance=7cm] (excluded2) {TAs Excluded (n=31): \\ Surrogate endpoints not used \\explicitly (n=29) \\ Committee papers unavailable (n=2)};
    \node[block1, below of=fulltext, node distance=3cm] (total) {TAs using surrogate endpoints (n=18)};
    % Draw Edges 
    \path[line] (records) -- (titles);
    \path[line] (titles) -- (fulltext);
    \path[line] (fulltext) -- (total);
    \path[line] (titles) -- (excluded1);
    \path[line] (fulltext) -- (excluded2);
\end{tikzpicture}
 \captionof{figure}{Flow Chart for Inclusion of Technology Appraisals} \label{fig:flowchart}
\end{center}

Of the 49 appraisals where there was a submission, 14 (29 percent) appraisals were in indications for blood and bone marrow cancers (including large B-cell lymphoma, multiple myeloma, Waldenstrom's macroglobulinaemia, acute myeloid leukaemia, chronic myeloid leukaemia, chronic lymphocytic leukaemia and Hodgkin's lymphoma) and the remaining 35 (71 percent) of appraisals were in solid tumours (including lung cancer, breast cancer, renal cancer, skin cancer, metastases and bladder cancer). For the 49 appraisals which were fully reviewed, the health technology being appraised was recommended for use in the NHS in 40/49 (82 percent) of appraisals and recommended for use in the Cancer Drugs Fund in 4/49 (8 percent) of appraisals. Table \ref{tab:char} summarises the characteristics of those 18 TAs where use of the surrogate endpoints was discussed explicitly. Of those 18 appraisals utilising surrogate endpoints the health technology being appraised was recommended for use in the NHS in 15/18 (83 percent) of appraisals and recommended for use in the Cancer Drugs Fund in 2/18 (11 percent) of appriasals. 

\begin{landscape}
\tiny
\begin{longtable}[c]{@{}llllllllllll@{}}
\caption{Characteristics of NICE technology appraisals using surrogate endpoints}
\label{tab:char}\\
\toprule
TA    & Pivotal Trial                                                            & Type of trial                                                                                             & \begin{tabular}[c]{@{}l@{}}No.\\ patients\end{tabular} & Treatment                                                                                                                  & Comparator                                                                                                                  & Disease                                                                                  & Surrogate & Final~ & \begin{tabular}[c]{@{}l@{}}Economic \\ Model\end{tabular}            & \begin{tabular}[c]{@{}l@{}}Time \\ Horizon\end{tabular}                   & TAR \\* \midrule
\endfirsthead
\multicolumn{12}{c}%
{{\bfseries Table \thetable\ continued from previous page}} \\
\toprule
TA    & Pivotal Trial                                                            & Type of trial                                                                                             & \begin{tabular}[c]{@{}l@{}}No.\\ patients\end{tabular} & Treatment                                                                                                                  & Comparator                                                                                                                  & Disease                                                                                  & Surrogate & Final~ & \begin{tabular}[c]{@{}l@{}}Economic \\ Model\end{tabular}            & \begin{tabular}[c]{@{}l@{}}Time \\ Horizon\end{tabular}                   & TAR \\* \midrule
\endhead
\bottomrule
\endfoot
\endlastfoot
TA885~\cite{TA885} & \begin{tabular}[c]{@{}l@{}}Keynote-826\\ (NCT03635567)\end{tabular}      & \begin{tabular}[c]{@{}l@{}}Phase 3, randomised, \\ double-blind, \\ placebo-controlled trial\end{tabular} & 617                                                    & \begin{tabular}[c]{@{}l@{}}Pembrolizumab and \\ chemotherapy with or \\ without Bevacizumab\end{tabular}                   & \begin{tabular}[c]{@{}l@{}}Placebo with chemotherapy \\ with or without Bevacizumab\end{tabular}                            & Cervical cancer                                                                          & PFS       & OS     & \begin{tabular}[c]{@{}l@{}}STM\\ (3 states)\end{tabular}             & 50 years                                                                  & CDF \\
TA876~\cite{TA876} & \begin{tabular}[c]{@{}l@{}}CheckMate-816\\ (NCT02998528)\end{tabular}    & \begin{tabular}[c]{@{}l@{}}Phase 3, randomised, \\ open-label, \\ active-controlled trial\end{tabular}    & 352                                                    & \begin{tabular}[c]{@{}l@{}}Nivolumab with \\ chemotherapy\end{tabular}                                                     & Chemotherapy                                                                                                                & NSCLC                                                                                    & pCR       & EFS    & \begin{tabular}[c]{@{}l@{}}STM\\ (4 states)\end{tabular}             & 35 years                                                                  & Yes \\
      &                                                                          &                                                                                                           &                                                        &                                                                                                                            &                                                                                                                             &                                                                                          &           & OS     &                                                                      &                                                                           &     \\
      &                                                                          &                                                                                                           &                                                        &                                                                                                                            &                                                                                                                             &                                                                                          & EFS       & OS     &                                                                      &                                                                           &     \\
TA874~\cite{TA874} & \begin{tabular}[c]{@{}l@{}}POLARIX\\ (NCT03274492)\end{tabular}          & \begin{tabular}[c]{@{}l@{}}Phase 3, randomised, \\ double-blind, \\ placebo-controlled trial\end{tabular} & 879                                                    & \begin{tabular}[c]{@{}l@{}}Polatuzumab vedotin, \\ rituximab, \\ cyclophosphamide, \\ doxorubicin and placebo\end{tabular} & \begin{tabular}[c]{@{}l@{}}Polatuzumab vedotin, \\ rituximab, \\ cyclophosphamide, \\ doxorubicin, vincristine\end{tabular} & \begin{tabular}[c]{@{}l@{}}Diffuse large \\ B-cell lymphoma\end{tabular}                 & PFS       & OS     & PSM                                                                  & 60 years                                                                  & Yes \\
      &                                                                          &                                                                                                           &                                                        &                                                                                                                            &                                                                                                                             &                                                                                          & EFS       & OS     &                                                                      &                                                                           &     \\
TA862~\cite{TA862} & \begin{tabular}[c]{@{}l@{}}DESTINY-Breast03\\ (NCT03529110)\end{tabular} & \begin{tabular}[c]{@{}l@{}}Phase 3, randomised, \\ open-label, \\ active-controlled trial\end{tabular}    & 524                                                    & Trastuzumab deruxtecan                                                                                                     & Trastuzumab emtansine                                                                                                       & Breast cancer                                                                            & PFS       & OS     & PSM                                                                  & 30 years                                                                  & Yes \\
TA851~\cite{TA851} & \begin{tabular}[c]{@{}l@{}}KEYNOTE-522\\ (NCT03036488)\end{tabular}      & \begin{tabular}[c]{@{}l@{}}Phase 3, randomised, \\ double-blind, \\ placebo-controlled trial\end{tabular} & 1174                                                   & \begin{tabular}[c]{@{}l@{}}Adjuvant or neoadjuvant \\ Pembrolizumab\end{tabular}                                           & Placebo                                                                                                                     & Breast cancer                                                                            & pCR       & EFS    & \begin{tabular}[c]{@{}l@{}}STM\\ (4 states)\end{tabular}             & 51 years                                                                  & Yes \\
      &                                                                          &                                                                                                           &                                                        &                                                                                                                            &                                                                                                                             &                                                                                          &           & OS     &                                                                      &                                                                           &     \\
      &                                                                          &                                                                                                           &                                                        &                                                                                                                            &                                                                                                                             &                                                                                          & EFS       & OS     &                                                                      &                                                                           &     \\
TA837~\cite{TA837} & \begin{tabular}[c]{@{}l@{}}KEYNOTE-716\\ (NCT03553836)\end{tabular}      & \begin{tabular}[c]{@{}l@{}}Phase 3, randomised, \\ double-blind, \\ placebo-controlled trial\end{tabular} & 954                                                    & Adjuvant Pembrolizumab                                                                                                     & Placebo                                                                                                                     & Melanoma                                                                                 & RFS       & DMFS   & \begin{tabular}[c]{@{}l@{}}STM\\ (4 states)\end{tabular}             & 40.7 years                                                                & Yes \\
      &                                                                          &                                                                                                           &                                                        &                                                                                                                            &                                                                                                                             &                                                                                          &           & OS     &                                                                      &                                                                           &     \\
      &                                                                          &                                                                                                           &                                                        &                                                                                                                            &                                                                                                                             &                                                                                          & DMFS      & OS     &                                                                      &                                                                           &     \\
TA830~\cite{TA830} & \begin{tabular}[c]{@{}l@{}}KEYNOTE-564\\ (NCT03142334)\end{tabular}      & \begin{tabular}[c]{@{}l@{}}Phase 3, randomised, \\ double-blind, \\ placebo-controlled trial\end{tabular} & 994                                                    & Adjuvant Pembrolizumab                                                                                                     & Placebo                                                                                                                     & \begin{tabular}[c]{@{}l@{}}Renal Cell \\ Carcinoma\end{tabular}                          & DFS       & OS     & \begin{tabular}[c]{@{}l@{}}STM\\ (4 states)\end{tabular}             & 41.1 years                                                                & Yes \\
TA833~\cite{TA833} & \begin{tabular}[c]{@{}l@{}}ASPEN\\ (NCT03053440)\end{tabular}            & \begin{tabular}[c]{@{}l@{}}Phase 3, randomised, \\ open-label, \\ active-controlled trial\end{tabular}    & 229                                                    & Zanubrutinib                                                                                                               & Ibrutinib                                                                                                                   & \begin{tabular}[c]{@{}l@{}}Waldenstrom's \\ Macroglobulinaemia\end{tabular}              & VGPR/CR   & PFS    & PSM                                                                  & 30 years                                                                  & Yes \\
      &                                                                          &                                                                                                           &                                                        &                                                                                                                            &                                                                                                                             &                                                                                          &           & OS     &                                                                      &                                                                           &     \\
TA823~\cite{TA823} & \begin{tabular}[c]{@{}l@{}}IMpower010\\ (NCT02486718)\end{tabular}       & \begin{tabular}[c]{@{}l@{}}Phase 3, randomised, \\ open-label, \\ active-controlled trial\end{tabular}    & 1280                                                   & Adjuvant Atezolizumab                                                                                                      & BSC and Chemotherapy                                                                                                        & NSCLC                                                                                    & DFS       & OS     & \begin{tabular}[c]{@{}l@{}}STM\\ (5 states)\end{tabular}             & 40 years                                                                  & CDF \\
TA817~\cite{TA817} & \begin{tabular}[c]{@{}l@{}}CheckMate274\\ (NCT02632409)\end{tabular}     & \begin{tabular}[c]{@{}l@{}}Phase 3, randomised, \\ double-blind, \\ placebo-controlled trial\end{tabular} & 709                                                    & Adjuvant Nivolumab                                                                                                         & Placebo                                                                                                                     & Urothelial cancer                                                                        & DFS       & OS     & \begin{tabular}[c]{@{}l@{}}STM\\ (4 states)\end{tabular}             & 40 years                                                                  & Yes \\
TA813~\cite{TA813} & \begin{tabular}[c]{@{}l@{}}ASCEMBL\\ (NCT03106779)\end{tabular}          & \begin{tabular}[c]{@{}l@{}}Phase 3, randomised, \\ open-label, \\ active-controlled trial\end{tabular}    & 233                                                    & Asciminib                                                                                                                  & Bosutinib                                                                                                                   & \begin{tabular}[c]{@{}l@{}}Chronic Myeloid \\ Leukaemia\end{tabular}                     & MMR       & PFS    & \begin{tabular}[c]{@{}l@{}}Cumulative \\ survival model\end{tabular} & 50 years                                                                  & Yes \\
      &                                                                          &                                                                                                           &                                                        &                                                                                                                            &                                                                                                                             &                                                                                          &           & OS     &                                                                      &                                                                           &     \\
      &                                                                          &                                                                                                           &                                                        &                                                                                                                            &                                                                                                                             &                                                                                          & CyR       & PFS    &                                                                      &                                                                           &     \\
      &                                                                          &                                                                                                           &                                                        &                                                                                                                            &                                                                                                                             &                                                                                          &           & OS     &                                                                      &                                                                           &     \\
      &                                                                          &                                                                                                           &                                                        &                                                                                                                            &                                                                                                                             &                                                                                          & CCyR      & PFS    &                                                                      &                                                                           &     \\
      &                                                                          &                                                                                                           &                                                        &                                                                                                                            &                                                                                                                             &                                                                                          &           & OS     &                                                                      &                                                                           &     \\
      &                                                                          &                                                                                                           &                                                        &                                                                                                                            &                                                                                                                             &                                                                                          & ToT       & PFS    &                                                                      &                                                                           &     \\
      &                                                                          &                                                                                                           &                                                        &                                                                                                                            &                                                                                                                             &                                                                                          &           & OS     &                                                                      &                                                                           &     \\
TA810~\cite{TA810} & \begin{tabular}[c]{@{}l@{}}monarchE\\ (NCT03155997)\end{tabular}         & \begin{tabular}[c]{@{}l@{}}Phase 3, randomised, \\ open-label, \\ active-controlled trial\end{tabular}    & 5637                                                   & \begin{tabular}[c]{@{}l@{}}Adjuvant Abemaciclib \\ with endocrine therapy\end{tabular}                                     & Endocrine therapy                                                                                                           & Breast cancer                                                                            & IDFS      & OS     & \begin{tabular}[c]{@{}l@{}}STM\\ (5 states)\end{tabular}             & 49 years                                                                  & Yes \\
      &                                                                          &                                                                                                           &                                                        &                                                                                                                            &                                                                                                                             &                                                                                          & DRFS      & OS     &                                                                      &                                                                           &     \\
TA796~\cite{TA796} & \begin{tabular}[c]{@{}l@{}}SACT data cohort study\\ (NA)\end{tabular}    & SACT data cohort study                                                                                    & 406                                                    & Venetoclax                                                                                                                 & NA                                                                                                                          & \begin{tabular}[c]{@{}l@{}}Chronic Lymphocytic \\ Leukaemia\end{tabular}                 & ToT       & PFS    & PSM                                                                  & 5, 10, 15 years                                                           & Yes \\
TA795~\cite{TA795} & \begin{tabular}[c]{@{}l@{}}SACT data cohort study\\ (NA)\end{tabular}    & SACT data cohort study                                                                                    & 823                                                    & Ibrutinib                                                                                                                  & NA                                                                                                                          & \begin{tabular}[c]{@{}l@{}}Waldenstrom's \\ Macroglobulinaemia\end{tabular}              & ToT       & PFS    & \begin{tabular}[c]{@{}l@{}}STM\\ (5 states)\end{tabular}             & \begin{tabular}[c]{@{}l@{}}Lifetime \\ (years not specified)\end{tabular} & Yes \\
TA784~\cite{TA784} & \begin{tabular}[c]{@{}l@{}}NOVA\\ (NCT01847274)\end{tabular}             & \begin{tabular}[c]{@{}l@{}}Phase 3, randomised, \\ double-blind, \\ placebo-controlled trial\end{tabular} & 553                                                    & Maintainance Niraparib                                                                                                     & Placebo                                                                                                                     & \begin{tabular}[c]{@{}l@{}}Ovarian, fallopian tube \\ and peritoneal cancer\end{tabular} & PFS       & OS     & \begin{tabular}[c]{@{}l@{}}Means based \\ model\end{tabular}         & 40 years                                                                  & Yes \\
TA772~\cite{TA772} & \begin{tabular}[c]{@{}l@{}}Keynote-204\\ (NCT02684292)\end{tabular}      & \begin{tabular}[c]{@{}l@{}}Phase 3, randomised, \\ open-label, \\ active-controlled trial\end{tabular}    & 304                                                    & Pembrolizumab                                                                                                              & Brentuximab Vedotin                                                                                                         & Hodgkins lymphoma                                                                        & MRD       & PFS    & PSM                                                                  & 50 years                                                                  & Yes \\
      &                                                                          &                                                                                                           &                                                        &                                                                                                                            &                                                                                                                             &                                                                                          &           & OS     &                                                                      &                                                                           &     \\
      &                                                                          &                                                                                                           &                                                        &                                                                                                                            &                                                                                                                             &                                                                                          & PFS       & OS     &                                                                      &                                                                           &     \\
TA763~\cite{TA763} & \begin{tabular}[c]{@{}l@{}}CASSIOPEIA\\ (NCT02541383)\end{tabular}       & \begin{tabular}[c]{@{}l@{}}Phase 3, randomised, \\ open-label, \\ active-controlled trial\end{tabular}    & 1085                                                   & \begin{tabular}[c]{@{}l@{}}Daratumumab and \\ Bortezomib\end{tabular}                                                      & Bortezomib                                                                                                                  & Multiple myeloma                                                                         & sCR       & PFS    & \begin{tabular}[c]{@{}l@{}}Response \\ based\end{tabular}            & 40 years                                                                  & Yes \\
      &                                                                          &                                                                                                           &                                                        &                                                                                                                            &                                                                                                                             &                                                                                          &           & OS     &                                                                      &                                                                           &     \\
TA766~\cite{TA766} & \begin{tabular}[c]{@{}l@{}}Keynote-054\\ (NCT02362594)\end{tabular}      & \begin{tabular}[c]{@{}l@{}}Phase 3, randomised, \\ double-blind, \\ placebo-controlled trial\end{tabular} & 1019                                                   & Adjuvant Pembrolizumab                                                                                                     & Placebo                                                                                                                     & Melanoma                                                                                 & RFS       & OS     & \begin{tabular}[c]{@{}l@{}}STM\\ (4 states)\end{tabular}             & 46 years                                                                  & Yes \\* \bottomrule
\end{longtable}

\noindent\tiny{TA = technology appraisal, NSCLC = non-small cell lung cancer, PFS = progression-free survival, pCR = pathologic complete response, EFS = event-free survival, RFS = recurrence-free survival, DFS = disease-free survival, VGPR = very good partial response, CR = complete response, MMR = major molecular response, CyR = cytogenic response, CCyR = complete cytogenic response, ToT = time on treatment, IDFS = invasive disease free survival, DRFS = distant relapse free survival, MRD = minimal residual disease, sCR = stringent complete response, OS = overall survival, DMFS = distant metastasis free survival, TAR = Technology Appraisal Recommendation, CDF = Cancer Drugs Fund, STM = state transition model, PSM = partitioned survival model, SACT = systemic anti-cancer therapy dataset}

\end{landscape}

\subsection{Rationale for and scope of use of surrogate endpoints and indications}

Surrogate endpoints were used in either the clinical-effectiveness or cost-effectiveness analyses of 18 technology appraisals (Table \ref{tab:char}). In the majority of appraisals (83 percent) it was stated that use of surrogate endpoints was required due to ``immature OS data", indicating that surrogate endpoints are primarily used when trials are ongoing, resulting in limited evidence on the final clinical outcome. 

There were three TAs where ``immature OS data" was not given as a reason for the use of surrogate endpoints. In two appraisals (TA795~\cite{TA795} and TA796~\cite{TA796}), time on treatment (ToT) was used as a proxy, or surrogate endpoint, for progression-free survival (PFS), as data on PFS were unavailable from the primary source of evidence for the submission. In the cancer drugs fund (CDF) appraisal TA784~\cite{TA784}, mature data on OS were collected from a pivotal trial conducted between the original submission and updated CDF review. However, despite the availability of mature OS data, TA784 utilises PFS as a surrogate endpoint for OS stating that mature OS data were difficult to interpret due to high levels of missing data and crossover between treatment arms. 

TAs utilising surrogate endpoints included many cancer types and the majority of indications (61 percent) were in solid tumours including breast cancer, lung cancer, skin cancer, cervical cancer, renal cancer, bladder cancer and ovarian cancer. The remaining 39 percent of indications were in blood and bone marrow cancers including Waldenstrom's macroglobulinaemia, large B-cell lymphoma, chronic myeloid leukaemia, Hodgkin's lymphoma, multiple myeloma and chronic lymphocytic leukaemia. 

\subsection{Types of outcome measures used as  surrogate endpoints}

Ten of the 18 (56 percent) technology appraisals discussed the use of a single surrogate endpoint (e.g. PFS as a surrogate endpoint for OS). However, the remaining 8 appraisals considered more than one surrogate endpoint in a single appraisal. For example TA876~\cite{TA876}, evaluating nivolumab with chemotherapy for the treatment of non-small cell lung cancer, discussed pathological complete response (pCR) as a surrogate endpoint to event-free survival (EFS) and OS in addition to EFS as a surrogate endpoint to OS. In total 37 surrogate endpoints were considered in the 18 technology appraisals. 

The most frequently considered surrogate endpoint was PFS, which was used in the evaluation of 5 (28 percent) technologies (pembrolizumab in two indications, polatuzumab vedotin, trastuzumab deruxtecan and niraparib). EFS and disease-free survival (DFS) were each used in the evaluation of 3 (17 percent) technologies (EFS: nivolumab, polatuzumab vedotin and pembrolizumab; DFS: pembrolizumab, atezolizumab and nivolumab). Recurrence-free survival (RFS) and pCR were each used to evaluate 2 (11 percent) technologies (RFS: pembrolizumab in two indications; pCR: nivolumab and pembrolizumab). ToT was used as a surrogate endpoint to PFS in two appraisals (venetoclax and ibrutinib). Other surrogate endpoints utilised in the technology appraisals were very good partial response/complete response (VGPR/CR), major molecular response rate (MMR), cytogenic response (CyR), complete cytogenic response (CCyR), invasive disease-free survival (IDFS), distant relapse-free survival (DRFS), minimal residual disease (MRD) and stringent complete response (sCR). 

\subsection{The level of evidence provided to validate surrogate endpoints}

According to Ciani et al~\cite{ciani2017time}, the first stage in assessing the validity of a surrogate endpoint is to evaluate the level of evidence provided for the surrogate relationship, where the lowest level of evidence is a biologically plausible relationship, the second level of evidence is a consistent association between the surrogate endpoint and final outcome observed in observational studies and the highest level of evidence is a relationship between treatment effects on the surrogate endpoint and final outcome observed in randomised controlled trials. 

There were 37 separate surrogate endpoints discussed in the 18 technology appraisals using surrogate endpoints. 
For 11 (30 percent)~\cite{TA876, TA862, TA837, TA823, TA817, TA763, TA766} of these surrogate endpoints, evidence from RCTs was provided to validate the relationship. Technology appraisals typically cited a systematic review and meta-analysis of these trials or studies. For 7 (19 percent)~\cite{TA784, TA813, TA795, TA763} of putative surrogate endpoints, evidence from observational studies was cited (see Section \ref{section:RWE}) and a further 12 (32 percent)~\cite{TA885, TA851, TA837, TA830, TA813, TA810, TA784, TA772} stated that the use of surrogate endpoints was based on clinical opinion. 7 (19 percent)~\cite{TA833, TA813, TA796} of the surrogate endpoints discussed did not provide any evidence supporting their use. 

\subsubsection{Association between the surrogate endpoint and the final clinical outcome}

Following the recommendations by Ciani et al~\cite{ciani2017time}, the second stage in assessing the validity of a surrogate endpoint is to evaluate the strength of association between the surrogate endpoint and final outcome. The strength of association is often evaluated through measures such as the individual-level correlation and $R^2$. When assessing this association it would be preferable to obtain individual participant data (IPD) from RCTs in order to estimate association at the trial-level and the individual-level. However, such data are not always available. 

For the 37 putative surrogate endpoints, a measure of association was explicitly reported in the TAs for 6 surrogate endpoints. Three of these were obtained from TA876~\cite{TA876} investigating pCR as a surrogate endpoint for EFS and OS and EFS as a surrogate for OS. For the surrogate relationship between pCR and EFS/OS, the TA states that “patients who have a pCR have improved EFS and OS outcomes versus patients who do not achieve a pCR”. The TA then goes on to state that patient-level EFS by pCR has a HR = 0.49 (95\% CI: 0.41, 0.60) and that patient-level OS by pCR has a HR = 0.49 (95\% CI: 0.42, 0.57). For the surrogate relationship between EFS and OS, the TA states that there is a weighted Pearson’s correlation coefficient of r=0.819 (95\% CI: 0.73, 0.92) between median EFS and median OS and a positive linear correlation of 0.864 (95\% CI: 0.81, 0.99) between logHRs on EFS and OS. Furthermore, the TA provided an estimate of trial-level association in the form of and $R^2$ value of $0.777$ obtained from a random-effects meta-regression. Additional measures of association were reported in this TA however some measures could not be extracted due to censoring of the committee papers. 

However, not all TAs referenced specific association measures between surrogate endpoints and final clinical outcomes. For those surrogate endpoints where no measure of association was reported, this was often due to lack of evidence available to support the surrogate relationship. For example, some technology appraisals were unable to provide a measure of association, as use of the surrogate endpoint was based on clinical opinion alone with no evidence from observational studies or RCTs. However, several technology appraisals made statements such as ``RFS appears to be a valid surrogate endpoint for OS" (TA766~\cite{TA766}) and cited a paper describing this relationship. While some of the papers cited provided a measure of association between the surrogate endpoint and final outcome (for example the paper cited in TA766 reported a correlation of 0.89 between RFS and OS), others simply cited trial reports which provided no measure of association between the surrogate endpoint and final outcome (TA817~\cite{TA817}). 

\subsubsection{The relationship between the treatment effects on the surrogate endpoint and final outcome}

Continuing to follow the recommendations set out by Ciani et al~\cite{ciani2017time}, the final stage of assessing the validity of a surrogate endpoint is to evaluate the relationship between the surrogate endpoint and final outcome, and in particular between the treatment effects on the two endpoints. This typically involves estimating (predicting) the expected treatment effect on the final clinical outcome and can include estimating the surrogate threshold effect (STE), which is the magnitude of treatment effect on the surrogate endpoint which would predict a statistically significant/clinically meaningful treatment effect on the final outcome. 

Of the 37 surrogate endpoints evaluated in 18 technology appraisals, the relationship between the surrogate endpoint and final outcome was only quantified within the TAs in 2 cases. In the committee papers for TA862~\cite{TA862} it is stated that ``a difference in median PFS of 5, 10, 15 and 20 months between an intervention and a comparator would be expected to translate into approximately 8.7, 17.4, 26.2 and 35.0 months' additional median OS for the intervention". The company go on to say that ``the 17.9-month increase in median PFS $\dots$ is expected to translate into a clinically significant OS advantage". In the second instance, in TA553 (original appraisal for CDF appraisal TA766~\cite{TA766}), an STE of 0.77 is reported, indicating that for studies with a HR $\leq 0.77$ for RFS, there is a meaningful treatment benefit predicted on OS. In TA553, the HR on RFS is 0.57, supporting the argument for a treatment effect on OS. 

While only two putative surrogate endpoints had the prediction explicitly reported in the TA, several more make references to papers which conduct surrogate threshold effect analysis. For example, TA823~\cite{TA823} references a paper by Mauguen et al~\cite{mauguen2013surrogate} which states a STE of 0.88 and suggests that ``cross-validation results confirmed the accurate prediction of the treatment effect on OS based on the effects observed on DFS". 

\subsubsection{Use of non-randomised evidence}

Traditionally, reimbursement decisions would be made based on data from a pivotal phase 3 RCT. Of the 47 TAs reviewed, 35 (74 percent) were supported by data from a pivotal phase 3 RCT. However, 10 TAs were supported by either a phase 1 or phase 2 study and 2 TAs were supported by real-world evidence (RWE). 

The two TAs, which were supported by RWE, were CDF appraisals TA795~\cite{TA795} of Ibrutinib monotherapy for treatment of Waldenstrom's Macroglobulinaemia (WM) and TA796~\cite{TA796} of Venetoclax for treating chronic lymphocytic leukaemia (CLL). These reviews both use the systemic anti-cancer therapy (SACT) dataset to support their submission. SACT is a dataset which collects information on anti-cancer therapies used across all NHS England Trusts, including those approved for use in the CDF. SACT was chosen as the primary source of evidence for these TAs as it was deemed more generalisable to practice in NHS England. The limitation of SACT is that it typically does not record information on PFS and thus cannot be used to inform clinical or cost-effectiveness estimates. However, ToT is collected within SACT and in both TA795 and TA796, ToT was used as a surrogate, or proxy, to PFS. 

Of the 10 TAs which were supported by phase 1 or 2 trials, 8 of these were also single-arm trials, including 3 first-in-human phase 1 single-arm trials. The use of single-arm trials in HTA poses a problem for assessment of clinical and cost-effectiveness as there is no comparator arm to use as a baseline to estimate relative treatment effects. Instead, studies which utilise single-arm trials as the pivotal study supporting the submission must use external evidence such as historical trials or RWE to estimate relative treatment effects. While none of the TAs reviewed contained both use of surrogate endpoints and a single-arm trial, the increasing use of both means that it is likely that in the future, we will see more submissions to NICE which are based on single-arm trials investigating the treatment effect on a surrogate endpoint. 

\subsection{Use of surrogate endpoints in cost-effectiveness analysis} \label{section:RWE}

An economic model was presented for each of the 18 technology appraisals using surrogate endpoints. The most commonly used economic models were state-transition models (STMs) which were used in 10 (56 percent) of the appraisals and partitioned survival models (PSMs), which were used in 5 (27 percent) of the appraisals. The majority of appraisals justified the use of STMs by stating that such models are better able to model the disease pathway. However, some also stated that in scenarios where survival data are immature, STMs are preferable to PSMs as overall survival can be modelled as a function of all other health states in the model rather than requiring extrapolation from immature data. Appraisals utilising the partitioned survival approach justified using this method as it (a) directly utilises survival data from pivotal trials, (b) is intuitive and (c) is frequently used in oncology appraisals. Other economic models utilised were a cumulative survival model (TA813~\cite{TA813}), surrogate survival model (TA813), means-based model (TA784~\cite{TA784}) and response-based model (TA763~\cite{TA763}). 

For the five technology appraisals which utilised PSMs (TA874~\cite{TA874}, TA862~\cite{TA862}, TA833~\cite{TA883}, TA796~\cite{TA796} and TA772~\cite{TA772}), all appraisals employed a PSM with three states of “progression free”, “progressed disease” and “dead”. In a PSM with these states, the value of each health state is determined using time-to-event data represented by the survival curves for PFS and OS for the treatment of interest. In three appraisals using PSMs (TA874, TA862 and TA833) the surrogate endpoint was not used to support analysis as long-term OS was obtained via extrapolation of immature OS data from the pivotal trial. The use of such immature data for extrapolation to inform the economic model was noted as a concern by the evidence review group in each appraisal. In TA772, OS data from the pivotal trial were too immature for extrapolation. Instead, the company estimated long-term OS by extrapolation of OS from an external study, assuming no treatment benefit on OS. Therefore, in the absence of mature OS data, TA772 assumes no treatment benefit on OS rather than utilizing information on the surrogate endpoint to estimate the treatment effect on OS. Finally, for TA796 long-term PFS was estimated via extrapolation of time on treatment (ToT) from the systemic anti-cancer therapy (SACT) real-world evidence dataset. The use of ToT as a proxy for PFS was necessitated by the lack of reporting of PFS in SACT. 

For the ten technology appraisals which utilised STMs (TA885~\cite{TA885}, TA876~\cite{TA876}, TA851~\cite{TA851}, TA837~\cite{TA837},
TA830~\cite{TA830}, TA823~\cite{TA823}, TA817~\cite{TA817}, TA810~\cite{TA810}, TA795~\cite{TA795}, TA766~\cite{TA766}), the number and definition of health states varied across the appraisals. However, in any STM patients move between different health states representing stages of a disease. How quickly patients move between these health states is defined by transition probabilities, which are generally estimated from time-to-event data in pivotal trial. For 6 appraisals (TA885, TA876, TA851, TA810 and TA766, TA817), the majority of transition probabilities were estimated directly from their pivotal trials. Therefore, despite all six TAs providing statements from both the company and ERG indicating that data on OS is immature, mortality events from pivotal trials were still used to inform the economic model and the surrogate relationship was not utilised. However, in three appraisals (TA837, TA830 and TA823) the number of mortality events was so low that real world evidence (RWE) and external data were used in place of the pivotal trial to extrapolate over the model horizon. Therefore, these three appraisals preferred to use external but more mature data to inform their economic model rather than estimate mortality based on surrogate endpoints measured in their pivotal trials. Finally, for TA795 transition probabilities from the progression-free state were informed by estimating a HR for time to treatment discontinuation (TTD) in SACT versus TTD in the Rory Morrison Register (RMR). This HR was applied to an exponential model fitted to the PFS data from the RMR to obtain a long-term estimate of PFS in SACT. Therefore, TTD was not used directly as a proxy for PFS in TA795, instead it was assumed that the relationship which holds for TTD between SACT and RMR also holds for PFS. The use of this approach was necessitated by the lack of reporting of PFS in SACT.  

Finally for TA813~\cite{TA813}, TA784~\cite{TA784} and TA763~\cite{TA763}, alternative cost-effectiveness models to PSM and STM were used. In TA813, the company used a cumulative survival approach which uses TTD as a surrogate for survival outcomes. However, the ERG was concerned about the lack of evidence to link TTD with survival and preferred a surrogate survival approach which modelled PFS as a function of cytogenic and haematological response, as they felt these responses as surrogates were supported by the literature. In TA784, the company used a mean based model which is comprised of three health states: “progression-free”, “progressed disease” and “death”. Like a PSM, it estimates long-term PFS and OS but calculates area under the curve (AUC) to obtain mean costs and utilities which are applied to the average time spent in each health state. For PFS in both arms and OS in the treatment arm, long-term survival is estimated via extrapolation of pivotal trial data. However, the company argue that OS for the control arm cannot be extrapolated due to confounding and instead use a 1:1 PFS:OS relationship to estimate OS for the control arm. The ERG criticised this approach citing a lack of consistent evidence for the PFS-OS surrogate relationship. In TA763 a response-based model used KM curves from the pivotal trial up to a landmark point and split those alive at this point by minimal residual disease (MRD) status. The company followed a 5 stage approach to estimate PFS and OS for control and treatment arms. First PFS and OS for MRD+ in control arm were extrapolated. Second, a meta-analysis estimated HRs reflecting the association between MRD status and PFS and OS. Third, the HRs were applied to extrapolated curves from the first stage. Fourth, HRs obtained from analysis of pivotal trial were applies to curves to obtain PFS and OS for treatment groups split by MRD status. Finally, survival curves were weighted by MRD status. Therefore, the surrogate relationship between MRD status and PFS and OS was utilised within this model.

\section{Discussion}

In this project we investigated the use of surrogate endpoints in 18 NICE technology appraisals in oncology out of 47 technology appraisals available for review. We found that 37 separate surrogate endpoints were discussed in a range of oncology indications including haematological and solid tumours. The majority of technology appraisals used surrogate endpoints as a result of immature OS data from pivotal trials. 

Based on a previously proposed three-step approach for validating surrogate endpoints~\cite{ciani2017time} we found that 11 (30 percent), 7 (19 percent) and 12 (32 percent) of the surrogate endpoints were supported by evidence from RCTs, observational data and biological plausibility, respectively. For only 6 (16 percent) of the putative surrogate endpoints, did the technology appraisal present a measure of association between the surrogate endpoint and final outcome.
For those TAs which did report a measure of association between the surrogate endpoint and final outcome, the correlations reported between treatment effects on the surrogate endpoint and final outcome were reasonably strong (0.40-0.95). However, while such associations might be acceptable to NICE, who do not provide thresholds for the recommended acceptability of a surrogate, they would likely be less acceptable to the HTA body in Germany, IQWiG, who recommend that the lower confidence interval for correlation, $R$, is greater than 0.85~\cite{institute2011validity}. Only 2 (5 percent) of the putative surrogate endpoints were further supported by a quantification of the expected treatment effect on the final outcome based on the observed treatment effect on the surrogate endpoint. It is possible that a lack of evidence from RCTs on specific treatments and disease areas contributes to the paucity of statistical evidence for surrogate relationships. However, methods are emerging to explore borrowing information across treatment classes, disease indications and from different sources of evidence (such as registry data) to improve validation of surrogate relationships where data from RCTs on the relevant disease area and treatment are sparse~\cite{papanikos2020bayesian, singh2023multiindication, wheaton2023using}

For NICE oncology technology appraisals using surrogate endpoints, STMs were the most commonly used in cost-effectiveness analysis, with several appraisals noting that where there is immature survival data, such models are preferable as OS does not need to be directly modelled. However, PSMs were also frequently used, sometimes utilising highly uncertain extrapolations of immature OS curves from pivotal trials. The majority of appraisals preferred to utilise immature OS data from pivotal trials, external trials or real world evidence rather than incorporate information on the relationship between surrogate endpoints and final outcomes in the economic model.  

There are several limitations of this project. First, the selection of technology appraisals is based on documents citing the terms “surrogate” or “surrogacy”. While this approach expedites the review process, it is possible that this decision resulted in excluding technology appraisals where there was non-explicit discussion of surrogate endpoints as there can be discrepancies in how patients, clinicians, regulators and health technology assessment experts define and describe a surrogate endpoint~\cite{manyara2023definitions, ciani2023framework}. A future, more comprehensive review, could consider inclusion of terms such as “intermediate endpoint” and “biomarker” to more fully encompass all discussions around surrogate endpoints. Second, this review only covered oncology drugs evaluated between February 2022 and May 2023. In the period of time since this review was conducted several more oncology products will have been reviewed by NICE and there may have been further development of methods or guidelines on the use of surrogate endpoints in HTA. Therefore, future reviews of more recent TAs may show a more rigorous approach to the use of data on surrogate endpoints. Finally, this review has potentially limited generalisability as it was only conducted within NICE which conducts decision making for England and Wales. However, while this review evaluated the use of surrogate endpoints in oncology technology appraisals conducted by NICE for England and Wales, we hope that the outcomes of this review are beneficial to health technology assessment agencies internationally. We hope that by highlighting the lack of standardisation of methods for evaluating surrogate endpoints in NICE technology appraisals that this will prompt other HTA bodies to investigate how surrogate endpoints are used within their own appraisals. Furthermore, we hope that this will foster further international collaboration of HTA bodies to work towards a more standardised approach to the use of surrogate endpoints in technology appraisals. 

Whilst the relatively newly updated NICE methods guide, recommending more explicit validation of surrogate endpoints, is still to gain more traction and future appraisals will indicate the uptake of methods and approaches the guide recommends, there is still need for further guidance and methods development to support appropriate and effective use of surrogate endpoints in HTA decision making. For example, throughout this project it was observed in several technology appraisals that while there was no meaningful discussion of surrogate endpoints, the pivotal trial used to support the appraisal did not include overall survival as a primary endpoint. This potentially suggests that further guidance is required to support appropriate reporting of surrogate endpoints in technology appraisals. A number of recent initiatives highlighted such need. Recent extensions of the CONSORT and SPIRIT statements to include appropriate reporting of use of surrogate endpoints and their justification~\cite{manyara2022protocol} will hopefully improve the transparency of the use of surrogate endpoints in clinical trials and in turn in HTA policy making. The International Society of Pharmacoeconomics and Outcomes Research (ISPOR) set up a Task Force currently developing  guidance on emerging good practices for surrogate endpoint evaluation and validation of the relationship between outcomes quantitatively informing HTA~\cite{ispor}, whilst NICE is currently working with international organisations to develop guidelines for incorporating surrogate endpoints in cost-effectiveness analyses~\cite{Bouvy_2023}. 

\section{Conclusions}

Over recent years there has been significant discussion in the literature about the increasing use of surrogate endpoints in supporting regulatory and reimbursement decisions. This review supports the statement that surrogate endpoints are frequently used in oncology technology appraisals in England and Wales. Despite the wide availability of statistical methods and guidance on how to appropriately validate surrogate endpoints, this review highlights that the use and validation of surrogate endpoints can vary dramatically between technology appraisals which leads to uncertainty in decision-making. To reduce such uncertainty, it is important that technology appraisals utilising surrogate endpoints provide clear evidence supporting (a) the association between the surrogate endpoint and final outcome, (b) the association between treatment effects on the surrogate endpoint and final outcome and (c) the prediction of the treatment effect on the final outcome based on the observed effect on the surrogate endpoint. Providing clear evidence to support the surrogate relationship will enable greater confidence in the clinical and cost-effectiveness estimates obtained by HTA bodies which in turn will allow more robust reimbursement decisions and enable patients' to access best available treatments. 

\subsection*{Acknowledgements}
This research was funded by the University of Leicester and Medical Research Council methodology research grant [MR/T025166/1]. 
The authors received additional support from the Leicester NIHR Biomedical Research Centre (BRC). The views expressed are those of the authors and not necessarily those of the NIHR or the Department of Health and Social Care.

\subsection*{Conflict of Interest Statement}
LW has no competing interests to declare. 
SB is a member of the NICE Decision Support Unit (DSU) and the NICE Guidelines Technical Support Unit (TSU). She has served as a paid consultant, providing methodological advice, to NICE, Roche, RTI Health Solutions and IQVIA and has received payments for educational events from Roche and the University of Bristol.

\newpage

\nocite{TA883, TA884, TA881, TA879, TA870, TA865, TA866, TA857, TA858, TA855, TA849, TA850, TA836, TA827, TA831, TA819, TA816, TA812, TA801, TA802, TA798, TA789, TA788, TA786, TA787, TA783, TA781, TA780, TA779, TA770, TA765, TA766}
\bibliographystyle{vancouver}
\bibliography{ref2.bib}

\end{document}